\documentclass[conference]{IEEEtran}
\usepackage{cite}
\usepackage{graphicx}
\usepackage{epstopdf}
\usepackage{amsmath}
\usepackage{flushend}
\usepackage{amsfonts}
\usepackage{amssymb,bm,upgreek}
\usepackage{nicefrac}

\begin{document}

\title{Active Attack on User Load Achieving Pilot Design in Massive MIMO Networks}

\author{\IEEEauthorblockN{Noman Akbar and Shihao Yan}
\IEEEauthorblockA{Research School of Engineering, Australian National University, Acton, ACT, 2601,
Australia}Emails: noman.akbar@anu.edu.au, shihao.yan@anu.edu.au}

\markboth{Submitted to IEEE Globecom Workshop 2017}{Akbar \MakeLowercase{\textit{et
al.}}: Active Attack on User Load Achieving Pilot Sequence Design}

\maketitle

\begin{abstract}
In this paper, we propose an active attacking strategy on a massive multiple-input multiple-output (MIMO) network, where the pilot sequences are obtained using the user load-achieving pilot sequence design. The user load-achieving design ensures that the signal-to-interference-plus-noise ratio (SINR) requirements of all the users in the massive MIMO networks are guaranteed even in the presence of pilot contamination. However, this design has some vulnerabilities, such as one known pilot sequence and the correlation among the pilot sequences, that may be exploited by active attackers. In this work, we first identify the potential vulnerabilities in the user load-achieving pilot sequence design and then, accordingly, develop an active attacking strategy on the network. In the proposed attacking strategy, the active attackers transmit known pilot sequences in the uplink training and artificial noise in the downlink data transmission. Our examination demonstrates that the per-cell user load region is significantly reduced by the proposed attacking strategy. As a result of the reduced per-cell user load region, the SINR requirements of all the users are no longer guaranteed in the presence of the active attackers. Specifically, for the worst affected users the SINR requirements may not be ensured even with infinite antennas at the base station.
\end{abstract}

\IEEEpeerreviewmaketitle

\section{Introduction}\label{sec:intro}

Massive multiple-input multiple-output (MIMO) technology is considered as one of the key enablers of the future fifth generation (5G) wireless networks. In a massive MIMO network, base stations (BSs) are equipped with hundreds of antennas. Massive MIMO provides a number of lucrative advantages over the conventional MIMO systems. One of these benefits is the increase in the spectral and energy efficiency \cite{Boccardi2014}. In addition, the use of massive MIMO technologies achieves a higher throughput and reliability \cite{Yang2015}. Another important advantage of massive MIMO is that the channels between BSs and users become increasingly orthogonal \cite{Akbar16} as the number of antennas at BSs increases, which leads to the fact that the interference in the network will be significantly reduced. Recent research in the context of massive MIMO focused on resolving some specific key issues that limit the performance of massive MIMO. Among these issues, pilot contamination is considered as the most severe performance degrading factor in massive MIMO networks \cite{Akbar16}.

Pilot contamination occurs when the number of users in a cell is larger than the number of orthogonal pilot sequences, i.e., when it is not possible to allocate orthogonal pilot sequences to all the users and thus the pilot sequences are reused in the network. Pilot contamination is a performance bottleneck in massive MIMO networks \cite{Marzetta2010,Larsson2014,Ashikhmin2012}, because it still exists even when the number of antennas at the BSs approaches infinity. As such, a lot of recent research works focused on mitigating or reducing the detrimental affects of pilot contamination in massive MIMO networks (e.g., \cite{Fernandes2013,Jose2011,Yin2013,Muller2014,Akbar16,Akbar16a}). The recent research in pilot contamination can be generally categorized into five groups: protocol based methods \cite{Fernandes2013}, precoding based methods \cite{Jose2011}, angle-of-arrival based methods \cite{Akbar16b}, blind methods \cite{Muller2014}, and pilot sequence design methods \cite{Akbar16,Akbar16a,Ulukus2001,Cotae2006,Viswanath1999}.

Recently, a user load-achieving pilot sequence design algorithm has been proposed for a multi-cell multi-user massive MIMO network \cite{Akbar16} and the thorough performance analysis of this algorithm has been conducted \cite{Akbar16a}. The key idea of the user load-achieving pilot design is to first determine the user load region of the network under pilot contamination. Then, the algorithm allocates pilot sequences for all users in a distributed manner, which requires very little BS cooperation. The algorithm also allocates the downlink transmit power for all the users at BSs, such that the signal-to-interference-plus-noise ratio (SINR) requirements of all the users in the network can be guaranteed. We note that the user load-achieving pilot sequence design \cite{Akbar16a} is also referred to as the user capacity-achieving pilot sequence design in \cite{Akbar16}. The main advantage of the user load-achieving pilot design is that it guarantees the SINR requirements for all the users in the network when some specific conditions are met. We would like to highlight that as long as the SINR requirements are within the per-cell user load region, the pilot sequence design and downlink transmit power allocation can ensure the SINR requirements of all the users. Otherwise, the pilot sequence design may not be feasible to guarantee the SINR requirements of all the users in the considered massive MIMO network.

In this paper, we propose a strategy for an active attacker who aims at exploiting the vulnerability in the user load-achieving pilot sequence design to degrade the performance of massive MIMO networks. In the proposed strategy, the attacker exploits the known properties of the user load-achieving pilot design to deliberately increase the pilot contamination in the uplink training phase. In addition, during the downlink transmission phase, the active attacker transmits artificial noise (AN) to increase the interference to each user in the network. Notably, the attack strategy carefully exploits the design of the user load-achieving pilot design and degrades its performance, such that the SINR requirements for all the users in the network are no longer guaranteed with a certain number of antennas at the BS. We recall that the goal of the user load-achieving pilot design is to ensure the SINR requirements for all the users in the network. This goal cannot be achieved in the the presence of the active attacker with the proposed attack strategy. The main contributions of this work are summarized as follows.
\begin{enumerate}
\item
We identify potential vulnerabilities in the user load-achieving pilot sequence design. As shown in this work, these vulnerabilities can be exploited by an active attacker to significantly degrade the performance of a massive MIMO network, such that the SINR requirements of all the users in the network cannot be guaranteed by the user load-achieving pilot sequence design.
\item
We propose an active attacking strategy on the user load-achieving pilot sequence design in massive MIMO networks. Our examination shows that the user load region achieved by the user load-achieving pilot sequence design  is significantly reduced by the active attacker, such that the diverse range of SINR requirements is no longer supported. Specifically, with the active attack the SINR requirements for some users cannot be guaranteed even with an infinite number of antennas at the BSs.
\end{enumerate}

\section{User Load-Achieving Pilot Sequence Design and Its Vulnerabilities}
For the sake of completeness, in this section we first present the process for the user load-achieving pilot sequence design \cite{Akbar16a,Akbar16} and then identify its vulnerabilities that can be exploited by an active attacker.

\subsection{User Load-Achieving Pilot Sequence Design}

In the user load-achieving pilot sequence design, the user load is defined as the number of users that can be simultaneously served in a pilot-contaminated massive MIMO network, such that the SINR requirement of each individual user is guaranteed. To this end, the user load-achieving pilot design first determines the user load region of the network and then designs the pilot sequences accordingly. The key benefit achieved by this design is that it is capable of ensuring a diverse range of the SINR requirements for all the users simultaneously in the pilot-contaminated massive MIMO network. A thorough comparison of the design with existing pilot sequence designs demonstrates that the user load-achieving design can achieve a larger user load region and support a greater and diverse range of the SINR requirements. In addition, the user load-achieving pilot design guarantees the SINR requirement of all the users in the network with a finite $N_{t}$, where $N_{t}$ is the number of antennas at each BS. Meanwhile, the existing pilot designs are unable to support such diverse SINR requirements even with an infinite $N_{t}$. We next present the vulnerabilities in the user load-achieving pilot sequence design, which can be exploited by an active attacker.

\subsection{Vulnerabilities in the User Load-Achieving Pilot Design}

We first find that the user load-achieving pilot sequence design always outputs the pilot sequence of the form $[1,0,\cdots,0]^T$ for one user in each cell. As such, when the length of the pilot sequence is known, the attacker can figure out the pilot sequence assigned to at least one user in each cell. Furthermore, we note that the user load-achieving pilot sequence design modifies the SINR requirements for all the users in the network such that the SINR requirements lie on the upper surface boundary of the user load region. This SINR modification ensures that the benefits offered by the large user load region of the user load-achieving pilot design are fully utilized. On the other side of the coin, this SINR modification introduces the potential vulnerability in the pilot design (i.e., the known sequence $[1,0,\cdots,0]^T$). Importantly, in the presence of active attackers, the SINR requirements may no longer remain inside the user load region. Consequently, the SINR requirements of all the users in the network will not be guaranteed. Another vulnerability in the user load-achieving design is that all the pilot sequences designed for the network are correlated with each other. As such, if the attacker even knows one pilot sequences in the network, it can potentially contaminate the channel estimates of all the users in the massive MIMO network.
We note that an attacker only needs to know two network parameters for successfully exploiting the user load-achieving pilot design, i.e., the length of the pilot sequence, and the information that user load-achieving pilot design is being used in the network. We highlight that these parameters are easy to obtain in any network. Throughout this paper, we assume that the attacker has knowledge of these network parameters.

\section{Multi-Cell Massive MIMO Networks with Active Attackers}\label{sec:system}

In this section, we first detail the adopted system model and related assumptions. Then, we present the channel estimation and data transmission in the presence of the active attackers.

\subsection{System Model and Adopted Assumptions}

\begin{figure}[!t]
\centering
\includegraphics[width=21pc]{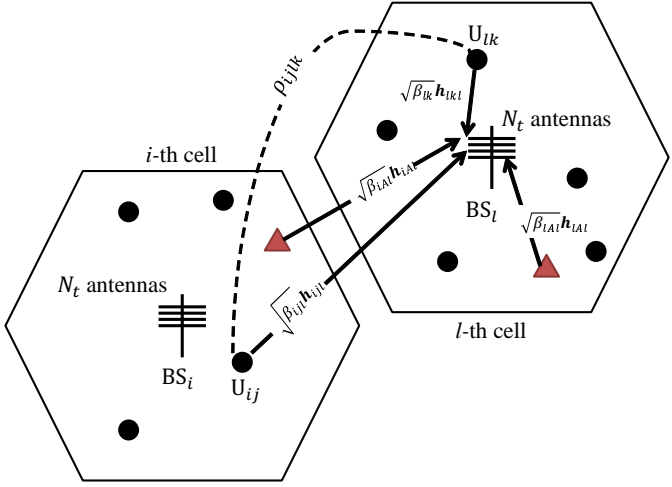}
\caption{An illustration of a multi-cell multi-user massive MIMO network in the presence of active attackers. Legitimate users are represented by black circles and active attackers are represented by red triangles. }
\label{network}
\end{figure}

We consider a multi-cell multi-user massive MIMO network, where there are $L$ cells and each of them has $K$ single-antenna users, as depicted in Fig.~\ref{network}. One BS is located in the center of each cell and is equipped with $N_{t}$ antennas. We assume that there is one active attacker present in each cell and thus totally there are $L$ active attackers in the network. We also assume that the communication channels in the network suffer from both large-scale and small-scale propagation effects. We denote the large-scale propagation factor from the $j$th user in the $i$th cell to the BS in the $l$th cell as $\beta_{i_{j}l}$. Additionally, we denote the small-scale propagation factor from the $j$th user in the $i$th cell to the $n$th BS antenna in the $l$th cell as $h_{i_{j}l_{n}}$. Consequently, the uplink propagation factor from the $j$th user in the $i$th cell to the $n$th BS antenna in the $l$th cell is represented as $\sqrt{\beta_{i_{j}l}}h_{i_{j}l_{n}}$. Furthermore, we assume the the small-scale propagation factor is Rayleigh distributed, i.e., $h_{i_{j}l_{n}}\sim\mathcal{CN}(0,1)$.
We assume that the network operates in the time-division-duplex (TDD) mode. The entire transmission, consisting of the uplink training and the downlink data transmission, occurs within one coherence block. As such, we assume that the uplink and the downlink channels remain unchanged for the entire transmission. As a result, the channel estimates in the uplink can be utilized for the downlink precoding \cite{Lu2014,Rusek2013}. We next detail the uplink channel estimation and downlink data transmission in the following two subsections.

\subsection{Channel Estimation with Active Attackers}

During the uplink training phase, the BS estimates the propagation factors from the users in a cell to the same-cell BS. Each user in a cell transmits a pre-assigned pilot sequence to the same-cell BS. We assume that all pilot sequences have unit energy. Additionally, the length of a pilot sequence is $\tau$, which is assumed to be the same for all the users in the network. In the presence of active attackers, the observation received at the $\textrm{BS}$ in the $l$th cell during the uplink training phase, denoted by $\mathbf{s}_{l}$, is given by
\begin{align}\label{rec_pilot}
\mathbf{s}_{l}=\sum_{i=1}^{L}\sum_{j=1}^{K}\eta_{i_{j}l}{\mathbf{Q}_{i_{j}}}\mathbf{h}_{i_{j}l}+\sum_{m=1}^L\eta_{m_{A}l}{\mathbf{Q}_{m_{A}}}\mathbf{h}_{m_{A}l} +\mathbf{n}_{l},
\end{align}
where $\eta_{i_{j}l} = \sqrt{p_{i_{j}}\beta_{i_{j}l}}$, $\mathbf{Q}_{i_{j}} = {\mathbf{q}_{i_{j}}}\otimes \mathbf{I}_{N_{t}}$ is the pilot matrix, $\mathbf{q}_{i_{j}}$ is the pilot sequence assigned to the $j$th user in the $i$th cell, $\otimes$ represents the Kronecker product, $\mathbf{I}_{N_{t}}$ denotes the $N_{t}\times{}N_{t}$ identity matrix, $p_{i_{j}}$ is the pilot power for the $j$th user in the $i$th cell, $\mathbf{h}_{i_{j}l}=[h_{i_{j}l_{1}},h_{i_{j}l_{2}},\dotsc,h_{i_{j}l_{n}}]^{T}$ is an $N_{t}\times1$ uplink channel vector from the $j$th user in the $i$th cell to the BS in the $l$th cell, and $\mathbf{n}_{l}$ is the additive white Gaussian noise (AWGN) at the BS in the $l$th cell. We highlight that the second term in the observation given in \eqref{rec_pilot} is due to the presence of the $L$ active attackers in the network.

Exploiting the structure of the pilot sequences generated by the user load-achieving pilot sequence design, we assume that all the attackers in the network transmit the known pilot sequence, i.e., $\mathbf{Q}_{m_{A}} = \mathbf{Q}_{1_{j}}$, where the pilot sequence assigned to $\mathbf{Q}_{1_{j}}$ is in the form $[1,0,\cdots,0]^T$.  Accordingly, the uplink channel from the $k$th user in the $l$th cell to the BS in the $l$th cell is obtained by utilizing the property of the pilot sequence matrix, given by $\mathbf{Q}_{l_{k}}^{T}\mathbf{Q}_{l_{k}}=\mathbf{I}_{N_{t}}$. Based on \eqref{rec_pilot} and assuming that the uplink power control is enabled with $\eta_{l_{k}l}=1$, the BS in the $l$th cell obtains the least square (LS) channel estimate for $\mathbf{h}_{l_{k}l}$ as $\mathbf{\hat{g}}_{l_{k}l}=\mathbf{Q}_{l_{k}}^{T}\mathbf{s}_{l}$. We rewrite $\mathbf{\hat{g}}_{l_{k}l}$ as
\begin{align}\label{channel_estimate_2}
\mathbf{\hat{g}}_{l_{k}l}\!=\! \mathbf{h}_{l_{k}l}\!+\! \sum_{i,j \neq l,k}\eta_{i_{j}l}\rho_{i_{j}l_{k}}\mathbf{h}_{i_{j}l}\!+\! \sum_{m=1}^L\eta_{m_{A}l}\rho_{m_{A}l_{k}}\mathbf{h}_{m_{A}l} \!+\! \bar{\mathbf{n}}_{l},
\end{align}
where $\mathbf{Q}_{l_{k}}^{T}$ denotes the matrix transpose of $\mathbf{Q}_{l_{k}}$, $\bar{\mathbf{n}}_{l}={\mathbf{Q}_{l_{k}}^T}\mathbf{n}_{l}$, ${\sum}_{i,j\neq{}l,k}=\sum_{i=1}^{L}\sum_{j=1}^{K}$ with the condition $(i,j)\neq(l,k)$, and $\rho_{i_{j}l_{k}}$ is the correlation coefficient between pilot sequences, defined as $\rho_{i_{j}l_{k}}=\mathbf{q}_{l_{k}}^{T}\mathbf{q}_{i_{j}}$, $k \in \left\{1,2,\dotsc, K\right\}$. We highlight that the uplink power control is not applied for the active attackers because the BSs are not aware of their presence.

In the user load-achieving pilot design, we have the value range of the correlation $\rho_{i_{j}l_{k}}$ in \eqref{channel_estimate_2} as  $ -1 \leq \rho_{i_{j}l_{k}} \leq +1$. If all the users are assigned orthogonal pilot sequences,  we have $\rho_{i_{j}l_{k}}=0$ and thus no pilot contamination. In massive MIMO networks, $\rho_{i_{j}l_{k}}$ is nonzero due to the limited number of orthogonal pilot sequences and thus  pilot contamination always exists. With the knowledge of pilot sequence assigned to one user in each cell, the active attackers can deteriorate the quality of the channel estimate by increasing pilot contamination, which is confirmed by the second term in \eqref{channel_estimate_2}. We note that the pilot sequences obtained from the user load-achieving design are correlated with each other. As such, the attacks not only affect the users with the pilot sequence of the form $[1,0,\cdots,0]^T$, but all the users in the network.

\subsection{Data Transmission via the Downlink}

We now focus on the downlink data transmission in the massive MIMO network. We assume that the attackers are active during this phase and transmit AN, while each BS transmits the downlink data symbols to the same-cell users. We denote the data symbol intended for the $k$th user in the $l$th cell as $x_{l_{k}}$. We also assume that the downlink transmit power for the symbol $x_{l_{k}}$ at the BS is given as $\mathbb{E}\left[x_{l_{k}}^H{x}_{l_{k}}\right] = P_{l_{k}}$, where $\mathbb{E}[\cdot]$ denotes the expectation operation. Based on the channel estimates obtained during the uplink training and the reciprocity between the uplink and downlink channels for the TDD mode, the BS performs a linear precoding using a vector $\mathbf{a}$. Thus, the received signal at the $k$th user in the $l$th cell is given by
\begin{align}\label{rec_initial}
\hat{y}_{l_{k}}=
\sum_{m=1}^{L}\sum_{n=1}^{K}\sqrt{\beta_{l_{k}m}}\mathbf{h}_{l_{k}m}^{H}\left(\mathbf{a}_{m_{n}}x_{m_{n}}\right)+w_{l_{k}},
\end{align}
where $w_{l_{k}}=\sum_{m=1}^L P_{m_{A}}w_{m_{A}} + \bar{w}_{l_{k}}$. Specifically, $w_{m_{A}}$ is the AN generated by the active attacker in the $m$th cell with transmit power $P_{m_{A}}$, and $\bar{w}_{l_{k}}$ is the AWGN at the $k$th user in the $l$th cell. \begin{figure*}[!t]
\begin{align}\label{long_exp}
\phi_{l_{k},N_{t}}=\frac{\left(\mathbb{E}\left[{\mathbf{h}_{l_{k}l}^{H}\mathbf{a}_{l_{k}}}\right]\right)^2\beta_{l_{k}l}P_{l_{k}}}
{\textrm{var}\left[{\mathbf{h}_{l_{k}l}^{H}\mathbf{a}_{l_{k}}}\right]\beta_{l_{k}l}P_{l_{k}} + \sum_{m,n{}\neq{}l,k}\mathbb{E}\left[|{\mathbf{h}_{l_{k}m}^{H}\mathbf{a}_{m_{n}}}|^{2}\right]\beta_{l_{k}m}P_{m_{n}} + \sigma_{w}^2}. \tag{5}
\end{align}
\begin{align}\label{SINR}
\phi_{l_{k},N_{t}}=\frac{\beta_{l_{k}l}P_{l_{k}}}{\left(\delta_{l_{k}}+\alpha_{l_{k}}\right)\left[\sum\limits_{m,n\neq{}l,k} \frac{\rho_{l_{k}m_{n}}^{2}\eta_{l_{k}m}^2\beta_{l_{k}m}P_{m_{n}}}{\left(\delta_{m_{n}}+ \alpha_{m_{n}}\right)}+\frac{1}{N_{t}}\left(\sum_{m=1}^{L}\sum_{n=1}^{K}\beta_{l_{k}m}P_{m_{n}} + \sigma_{w}^2\right)
\right]}. \tag{11}
\end{align}
\end{figure*}
Assuming that users only have the statistical information of the channel \cite{Jose2011,Shen2015}, we rewrite $\hat{y}_{l_{k}}$ in \eqref{rec_initial} as
\begin{align}\label{received_siga}
\hat{y}_{l_{k}}&\!=\!\sqrt{\beta_{l_{k}l}}\mathbb{E}\left[\mathbf{h}_{l_{k}l}^H\mathbf{a}_{l_{k}}\right]x_{l_{k}} \!+\!\sqrt{\beta_{l_{k}l}}\left({\mathbf{h}_{l_{k}l}^H\mathbf{a}_{l_{k}}}
\!\!-\!\!\mathbb{E}\left[{\mathbf{h}_{l_{k}l}^H\mathbf{a}_{l_{k}}}\right]\right)x_{l_{k}} \notag \\
&+ \sum_{m,n \neq l,k}\sqrt{\beta_{l_{k}m}}\mathbf{h}_{l_{k}m}^{H}\left(\mathbf{a}_{m_{n}}x_{m_{n}}\right)+w_{l_{k}}. \tag{4}
\end{align}

We now present the expressions for the achievable SINR for the $k$th user in the $l$th cell. We denote the achievable downlink SINR at the $k$th user in the $l$th cell by $\phi_{l_{k},N_{t}}$. We note that the first term in \eqref{received_siga} represents the signal intended for the $k$th user in the $l$th cell. We assume that the remanding terms in \eqref{received_siga} are uncorrelated with the intended signal and are treated as the effective noise. Accordingly, we express SINR $\phi_{l_{k},N_{t}}$ as \eqref{long_exp} given on the top of the page,
where $\textrm{var}\left[\cdot\right]$ denotes the variance operation, and $\sigma_{w}^2$ denotes the variance of $w_{l_{k}}$. We note that the SINR expression \eqref{long_exp} is a generalised expression valid for any type of linear precoding vector $\mathbf{a}_{l_{k}}$. Notably, we observe that the linear precoding vector is based on the channel estimates obtained by the uplink training phase. As such, pilot contamination in the uplink training affects the downlink data transmission.

In this work, we consider that the BS performs maximum-ratio transmission (MRT) precoding \cite{Akbar16,Shen2015}, which is given by
\begin{align}\label{MRT_precoder}
\mathbf{a}_{l_{k}}&=\frac{\mathbf{\hat{g}}_{l_{k}l}}{\|\mathbf{\hat{g}}_{l_{k}l}\|}=
\frac{\mathbf{\hat{g}}_{l_{k}l}}{\sqrt{N_{t}
\left(\mathbf{\hat{g}}_{l_{k}l}^{H}\mathbf{\hat{g}}_{l_{k}l}/N_{t}\right)}}, \tag{6}
\end{align}
where $\|\cdot\|$ denotes the $l_2$ norm. We now simplify the denominator in \eqref{MRT_precoder} by utilizing the fact that the channels in massive MIMO become increasingly orthogonal when the number of antennas at the BS (i.e., $N_t$) increases. This phenomenon is known as channel hardening and is represented as
\begin{align}\label{ortho_mm}
\frac{1}{N_t}\mathbf{h}_{i_{j}i}^{H}\mathbf{h}_{l_{k}l}=
\begin{cases}
1, & \forall~(i,j)=(l,k)\\
0, & \text{otherwise.} \tag{7}
\end{cases}
\end{align}
Using \eqref{ortho_mm}, we now simplify the denominator in \eqref{MRT_precoder} as
\begin{align}\label{MRT_precoder2}
\frac{\mathbf{\hat{g}}_{l_{k}l}^{H}\mathbf{\hat{g}}_{l_{k}l}}{N_t} &= \sum_{i=1}^{L}\sum_{j=1}^{K}\eta_{i_{j}l}^2\rho_{i_{j}l_{k}}^{2}+\sum_{m=1}^L\eta_{m_{A}l}^2\rho_{m_{A}l_{k}}^{2}+\sigma_{n_{l}}^{2}  \notag \\
&=\left(\delta_{l_{k}}+\alpha_{l_{k}}\right), \tag{8}
\end{align}
where $\alpha_{l_{k}} = \sum_{m=1}^L \eta_{m_{A}l}^2\rho_{m_{A}l_{k}}^{2}$ and
\begin{align}\label{delta_original}
\delta_{l_{k}} = \sum_{i=1}^{L}\sum_{j=1}^{K}\eta_{i_{j}l}^2\rho_{i_{j}l_{k}}^{2}+\sigma_{n_{l}}^{2}. \tag{9}
\end{align}
Substituting \eqref{MRT_precoder2} into \eqref{MRT_precoder}, we obtain the precoding vector as
\begin{align}\label{MRT_precoder3}
\mathbf{a}_{l_{k}}&=
\frac{\mathbf{\hat{g}}_{l_{k}l}}{\sqrt{N_{t}
\left(\delta_{l_{k}}+\alpha_{l_{k}}\right)}}. \tag{10}
\end{align}
We highlight that $\alpha_{l_{k}}$ in \eqref{MRT_precoder3} appears due to the pilot contamination caused by the active attackers. As such, the channel estimate $\mathbf{\hat{g}}_{l_{k}l}$ suffers from increased pilot contamination in the presence of the active attackers. We next present the closed form expression for $\phi_{l_{k},N_{t}}$, when the BSs adopt the precoding vector given in \eqref{MRT_precoder3} and the channel estimates are obtained using the LS channel estimation given in \eqref{channel_estimate_2}.
\newtheorem{lemma}{Lemma}
\begin{lemma}\label{lemma_red}
When the BSs adopt the precoding vector given in \eqref{MRT_precoder3} and the channel estimates are obtained using the LS channel estimation given in \eqref{channel_estimate_2}, the SINR at $k$th user in the $l$th cell is given in \eqref{SINR} at the top of the page.
\end{lemma}

In massive MIMO, each BS is equipped with a large number of antennas. We next present the asymptotic SINR when $N_{t}\rightarrow\infty$.  Following \eqref{SINR}, as $N_{t}\rightarrow\infty$ the asymptotic SINR expression for $\phi_{l_{k},N_{t}}$, denoted by $\phi_{l_{k},\infty}$, is given by
\begin{align}\label{SINR_infa}
\phi_{l_{k},\infty}\!=\!\frac{\beta_{l_{k}l}P_{l_{k}}}{\left(\delta_{l_{k}}\!+\!\alpha_{l_{k}}\right)\left(\sum\limits_{m=1}^{L}
\sum\limits_{n=1}^{K}\frac{\rho_{l_{k}m_{n}}^{2}\eta_{l_{k}m}^2\beta_{l_{k}m}P_{m_{n}}}{\left(\delta_{m_{n}}+\alpha_{m_{n}}\right)}\right)\!-\!\beta_{l_{k}l}P_{l_{k}}}. \tag{12}
\end{align}
We obtain some interesting observations from the the asymptotic SINR expression given by \eqref{SINR_infa}. The expression reveals that the pilot contamination still exists and limits the performance of massive MIMO, even when each BS is equipped with an infinite number of antennas. Furthermore, the increased pilot contamination due to the active attackers does not disappear in massive MIMO regime, i.e., $\alpha_{m_{n}}$ still exists when $N_{t}\rightarrow\infty$.

\section{User Load Region in Massive MIMO Networks with Active Attackers}

In this section, we present the user load region of the massive MIMO network in the presence of the active attackers, while the user load region without the active attackers is provided as a benchmark.

The user load-achieving pilot sequence design \cite{Akbar16,Akbar16a} guarantees the SINR requirements of all the users under the user load region. For comparison, we first represent the per-cell user load region of the massive MIMO network without active attackers, which is given by
\begin{align}\label{BW_all}
\sum_{i=1}^{L}\sum_{j=1}^{K}\left(\frac{\gamma_{i_{j}}}{1+\gamma_{i_{j}}}\right)\leq \frac{\tau}{L}, \tag{13}
\end{align}
where $\gamma_{i_{j}}$ is the SINR requirement of the $i$th user in the $j$th cell. Furthermore, ${\gamma_{i_{j}}}/{(1+\gamma_{i_{j}})}$ denotes the effective bandwidth of the $i$th user in the $j$th cell.
The bound on the user load signifies the region under which the user load is achieved, which means that the SINR requirements of all the users in the network  are guaranteed.

We now examine the impact of the active attackers on the user load region. As evident from \eqref{SINR}, the active attacks in the network lead to the reduction in the achievable SINR. The derivations for user load region with the active attackers are omitted here due to space limitations.  Following the similar approach as given in \cite{Akbar16,Akbar16a}, the per-cell user load region in the presence of the active attackers is given by
\begin{align}\label{BW}
\sum_{j=1}^{K}\left(\frac{\gamma_{i_{j}}}{1+\gamma_{i_{j}}}\right) + \left(\frac{\gamma_{m_{A}}}{1+\gamma_{m_{A}}}\right) \leq\frac{\tau}{L}, \tag{14}
\end{align}
where ${\gamma_{m_{A}}}/({1+\gamma_{m_{A}}})$ denotes the effective bandwidth of the active attacker in the $m$th cell.

We note that having active attackers in the network results in the reduction of the user load region. The BSs deign the user-load achieving pilot sequences based on the load region given in \eqref{BW_all}. We also note that the SINR requirements of all the users in the network can only be guaranteed within the user load region given in \eqref{BW_all}. With the active attackers, the user load region of the network is reduced to the one given in \eqref{BW}. As such, the SINR requirements of all the users in the network may not be supported by the BSs in the presence of the active attackers.

\section{Numerical Results}\label{sec:Numerical}

In this section, we numerically evaluate the proposed active attack strategy and compare the performance of the massive MIMO network with and without the active attackers. Specifically, we present numerical results to demonstrate the  performance degradation cased by the active attackers. Throughout this section, we consider a two-cell massive MIMO network, i.e., $L=2$. In addition, we set that there are eight users in the network and four users in each cell, i.e., $K=4$. Furthermore, the length of the pilot sequence used during the channel estimation is $3$, i.e., $\tau=3$. Each BS designs the pilot sequences based on the user load-achieving pilot sequence design proposed in \cite{Akbar16,Akbar16a}. Additionally, the downlink power for all the users in the network is set according to the pilot design \cite{Akbar16,Akbar16a}, where $P_{l_{k}}=\frac{\delta_{l_{k}}\gamma_{l_{k}}}{1+\gamma_{l_{k}}}$. Throughout this section, we assume that during the downlink transmission phase the active attackers transmit the AN with unit power, i.e., $P_{m_A}=1$.

\begin{figure}[!t]
\centering
\includegraphics[width=21pc]{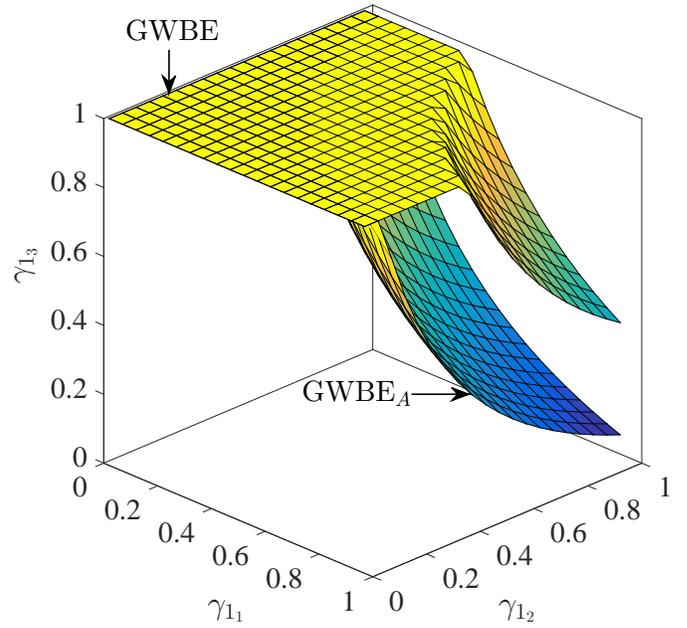}
\caption{The upper surface boundary of the per-cell user load regions versus the SINR requirements for the user load-achieving design with and without an active attacker.}
\label{capacity_region}
\end{figure}

We first compare the user load region of the network with and without the active attackers. In this comparison, we assume that the BS in each cell designs the pilot sequences separately. Accordingly, we compare the per-cell user load region. The SINR requirements for the users in the network are set as $\pmb{\gamma}_{1}=\pmb{\gamma}_{2}=\left[\gamma_{1_{1}},\gamma_{1_{2}},\gamma_{1_{3}},0.3\right]$. Additionally, we set that the effective bandwidth of the active attacker is 0.4. Fig.\ref{capacity_region} depicts the upper surface boundary of the per-cell user load region with and without the active attackers in the network. In this figure, the surface for the per-cell user load region without the active attackers, labeled as GWBE, is obtained using \eqref{BW_all} and the surface for the per-cell user load region with the active attackers, labeled as $\textrm{GWBE}_A$, is obtained using \eqref{BW}. We note that the user load region is significantly reduced by the active attackers, even with only one attacker in a cell. Specifically, having one active attacker in each cell reduces the user load region by approximately 24.69\%. The reduction in the user load region indicates that a group of users with high SINR requirements can no longer be successfully served in the pilot contaminated massive MIMO network in the presence of the active attackers.

\begin{figure}[!t]
\centering
\includegraphics[width=21pc]{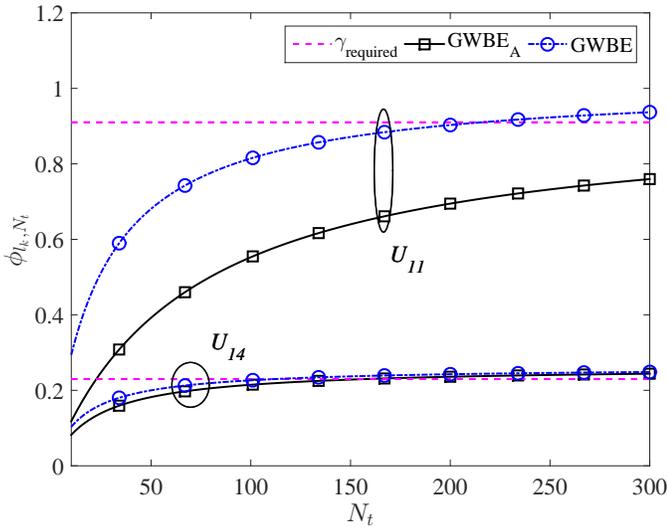}
\caption{The achievable SINR versus the number of antennas for the first user ($U_{11}$) and the fourth user ($U_{14}$) in the first cell for the user load-achieving design with and without an active attacker.}
\label{achievable SINR}
\end{figure}

In Fig.~\ref{achievable SINR}, we present the achievable downlink SINR performance with a finite number of antennas at the BSs with and without the active attackers. In this comparison, we generate results using \eqref{SINR} and consider that $L=2$, $\sigma_{w}^{2}=p_{l_{k}}=1$, and $\beta_{l_{k}m}=1$, where $l=m$, $\beta_{l_{k}m}=0.95$, and $l\neq m$. The SINR requirements for the users in the two cells are set as $\pmb{\gamma}_{1}=\left[0.91, 0.74, 0.64, 0.23\right], \pmb{\gamma}_{2}=\left[0.94, 0.82, 0.45, 0.10\right]$. We note that the SINR requirements are carefully selected such that they remain inside the user load reagin of the network depicted as GWBE in Fig.~\ref{capacity_region}. Additionally, we assume that the active attacker in each cell transmits the pilot sequence assigned to the first user in each cell. We clarify that the active attacker does not need to design the pilot sequences or know the full pilot sequence set designed for the entire network. Instead, the knowledge that the user load-achieving design is being used in the network is sufficient for the attacker to know the pilot sequences assigned to at least $L$ users in the network. Fig.~\ref{achievable SINR} depicts the achievable SINR for the two users in the first cell with and without one active attacker in the cell. We highlight that without the active attacker in the cell, the SINR targets of all the users in the network can be guaranteed. However, in the presence of the active attacker in the cell, the SINR requirements of all the users in the cell cannot be satisfied. For example, we observe that in the presence of the active attacker, the achievable SINR for the first user in the cell reduces from 0.90 to 0.69 when $N_t=200$. As such, there is an approximately $23.07\%$ reduction in the achievable SINR for this user. Importantly, the achievable SINR never meets the SINR requirement even when the number of antennas at the BS is infinite. In other words, the first user never achieves the SINR target in the presence of the active attacker. We note that the pilot sequence used by the first user in the network is the same as the one used by the active attacker. One important observation found  in Fig.~\ref{achievable SINR} is that the impact of the active attacker on the fourth user (i.e., $U_{14}$) in the cell is negligible. This is due to the fact that the correlation between the pilot sequence used by the fourth user and the pilot sequence adopted by the attacker is very small.

\section{Conclusions}\label{sec:con}

In this paper, we proposed an active attacking strategy on the user load-achieving pilot sequence design in a massive MIMO network. To this end, we first identified the potential vulnerabilities in the user load achieving pilot sequence design. Then, we proposed to increase the pilot contamination through transmitting known pilots by the active attackers in the uplink training. We demonstrated that the SINR requirements of all the users in the network are no longer guaranteed in the presence of the proposed active attacks. This is due to the fact that the per-cell user load region is significantly reduced by the active attackers. Specifically, the SINR requirements of the worst affected users by the attack may not be satisfied even with an infinite number of antennas at each BS in the massive MIMO network.

\section*{Acknowledgments}

This work was supported by the Australian Research Council's Discovery Projects (DP150103905) and the Australian Government Research Training Program (RTP) Scholarship.

\end{document}